\documentstyle[manuscript,aps,version2]{revtex}
\begin{document}
\draft

\title
\bf  

Transport in Sand Piles, Interface Depinning,\\
 and Earthquake Models
\endtitle

\author{Maya Paczuski$^1$ and Stefan Boettcher$^{2}$}
\instit
$^1$Department of Physics, Brookhaven National Laboratory,
Upton, NY 11973\\
$^2$Department of Physics and Astronomy, The
University of Oklahoma, Norman, OK 73019-0225
\endinstit

\medskip
\centerline{March 15, 1996. (Submitted to Phys. Rev. Lett.)}
\medskip

\abstract 

Recent numerical results for a model describing dispersive transport
in rice piles are explained by mapping the model to the depinning
transition of an interface that is dragged at one end through a random
medium.  The average velocity of transport vanishes with system size
$L$ as $<v> \sim L^{2-D}\sim L^{-0.23}$, and the avalanche size
distribution exponent $\tau= 2 - 1/D \simeq 1.55$, where $D\simeq
2.23$ from interface depinning.  We conjecture that the purely deterministic
Burridge-Knopoff ``train'' model for earthquakes is in the same
universality class.

\endabstract


Evidence for self-organized criticality \cite{soc} has been found in
controlled experiments on the granular dynamics of rice piles
\cite{frette}.  By slowly adding elongated rice grains in a narrow gap
between two clear plates, Frette {\it et al} found that the rice pile
evolves to a stationary angle of repose where on average one grain of
rice falls off the edge for every grain added at the wall.  Thereafter
transport of the rice through the pile occurs in terms of bursts with
no characteristic scale other than the system size.  The rice pile
exhibits self-organized criticality (SOC).  Subsequently, the Oslo
group has investigated tracer dispersion in the SOC pile by coloring
rice grains.  Christensen {\it et al} \cite{christ} found that the
average transport velocity of rice vanishes as the system size
diverges, and correspondingly that the distribution of transit times
through the pile was broad.  They proposed a ``sand pile'' model,
herein referred to as the Oslo model, to phenomenologically describe
their experiments on one-dimensional rice piles.

We establish that a broad universality class exists for SOC phenomena.
The Oslo ``sand pile'' model is mapped exactly to a model for
interface depinning where the interface is slowly pulled at one end
through a medium with quenched random pinning forces.  The height of
the interface maps to the number of toppling events in the sand pile
model.  The annealed noise of the random thresholds for toppling sand
grains maps to quenched pinning forces for the interface.  Thus a
problem of dispersive transport \cite{dispersive} in a granular medium
can be recast in terms of the somewhat better understood problem of
interface depinning \cite{scaling}.  This leads to a number of scaling
relations expressing critical exponents in the Oslo sand pile model in
terms of the avalanche dimension $D$.  This quantity is equal to the
avalanche dimension for a uniformly driven interface, which has been
determined numerically to be $D=2.23 \pm 0.03$ \cite{scaling}.  We also
predict that the avalanche size distribution exponent $\tau= 2 - 1/D
\simeq 1.55$ in the Oslo model, 
and that the average velocity of
transport vanishes as $<v> \sim L^{2-D}\sim L^{-0.23}$ when time is
measured in units of the number of sand grains added.  These
predictions agree with previous numerical simulation results
\cite{christ}, and with our simulation results.
 Finally, we conjecture that the Burridge-Knopoff
\cite{bk} train model studied by de Sousa Vieira \cite{maria}, a
purely deterministic mechanical model with no embedded randomness of a
spring-block chain pulled at one end, is also in the same universality
class.

The Oslo sand pile model is defined as follows: In a one dimensional
system of size $L$, an integer variable $h(x)$ gives the height of the
pile at position $x$, and $z(x)=h(x) -h(x+1)$ is the local slope.  The
boundary condition is $h(L+1)=0$.  Grains are dropped at $x=1$ until
the slope $z_1 >z_1^c$; then the site topples and one grain is
transferred to the neighboring site on the right $x=2$.  At each
subsequent time step, all sites $x$ with $z_x > z_x^c$ topple in
parallel.  In a toppling event at site $x$, $h(x)\rightarrow h(x)-1$
and $h(x+1) \rightarrow h(x+1) + 1$.  No grains are added to the pile
until the avalanche resulting from adding a sand grain ends and
the system reaches a stable
state with $z_x < z_x^c$ for all $x$.  The key ingredient making this model
different from previous sand pile models \cite{soc,limited} is that the
critical slopes $z_x^c$ are dynamical variables chosen randomly to be
1 or 2 every time a site topples.  The annealed randomness describes
in a simple way the changes in the local slopes observed in the rice
pile experiments \cite{christ}.

It is useful to define a local force
\begin{equation}
F(x,t) = h(x,t) - h(x+1,t) - \eta(x,H) \quad , \quad  1\leq x \leq L \quad ,
\end{equation}
where $\eta$ are the randomly distributed critical slopes, i.e.
$\eta(x,H) = z_x^c$ which take integer values 1 or 2 with equal
probability.  The boundary condition is $h(L+1,t)=0$ for all times.
At each time step $t \rightarrow t+1$, all unstable sites where
$F(x,t) >0$ topple.  The quantity $H(x,t)$ in Eq. (1) is the total
number of toppling events at site $x$ up to time $t$.  The threshold
slope at a site is chosen randomly after each toppling event at that
site; hence $\eta(x,H)$ is an uncorrelated quenched random variable in
the space of $(x,H)$.  The dynamics is central seeding; when all sites
have reached a stable state where $F(x) \leq 0$, a grain of sand is
added at site 1, $h(1) \rightarrow h(1) +1$, $t \rightarrow t+1$, and
a new avalanche starts.  It is easy to see that the sand pile dynamics
of toppling events is traced out by an advancing interface where the
height profile of the interface, $H(x,t)$, is the accumulated number
of topplings at that site.

Starting with an empty sand pile, $h(x,t=0)=0$ for all $x$, the number of
sand grains at $x$ at time $t$ is the local gradient in the number of
topplings that have occurred up to that time;
\begin{equation}
h(x,t) = H(x-1,t) - H(x,t) \quad .
\end{equation}
As a result, Eq. (1) can be rewritten as a dynamical equation for an 
interface with height profile $H(x,t)$,
\begin{equation}
F(x,t) = \nabla^2 H(x,t) - \eta(x,H) \quad , \ \quad 1\leq
x \leq L \quad ,
\end{equation}
 where the discretized Laplacian $\nabla^2 H(x)= H(x-1) -2H(x) +
H(x+1)$.  The interface dynamics is that for all $x$ where $F(x,t) >
0$, $H(x,t+1) \rightarrow H(x,t) + 1$ and the site advances; otherwise
$H(x,t+1)=H(x,t)$ and the site is pinned.  The boundary condition is
$H(L+1,t)= H(L,t)$ for all $t$.  Whenever the interface becomes stuck
so that $F(x) \leq 0$ for all $x$, it is pulled at the boundary at the
origin, $H(0) \rightarrow H(0) +1$.  This is an example of interface
depinning which has been widely studied
\cite{fegelman,leschhorn,scaling}.  The difference here is that rather
than being driven uniformly, the interface is driven by being slowly
dragged at the boundary.  After sufficient amount of motion has
occurred, the interface approaches a self-organized critical depinning
transition.  In the critical state, information about the pull at one
end can be communicated throughout the entire length of the interface.
This occurs when the interface finds an average curvature which
precisely balances the pinning forces.  This corresponds to the sand
pile reaching its critical angle of repose, where it can transfer sand
out from the origin to the edge of the pile.

In order to proceed, we briefly review some known results for
interface depinning with uniform driving \cite{scaling}.  The
depinning transition can either be reached by applying a constant
force or constant velocity constraint.  In constant force depinning a
uniform force is applied to all sites, i.e. add a term $F_{ext}$ to
the right hand side of Eq. (3), and advance all unstable sites with $F
>0$ in parallel.  When $F_{ext}$ is tuned to $F_c$ a depinning
transition occurs.  The constant velocity depinning transition is an
attractor for an extremal dynamics where the unique site along the
interface with the largest force $F(x,t)$ is advanced.  Now the
dynamics occurs in series with one site advancing at each step rather
than in parallel.  The extremal model self-organizes to the critical
state of constant velocity depinning.  The critical exponents relating
to the physical extents of completed avalanches are the same in the
constant force and constant velocity cases; i.e the roughness exponent
$\chi$, the avalanche dimension $D$, the exponent for the distribution
of avalanche sizes $\tau$.  Since the dynamics is different, though,
the critical exponents referring to propagating avalanches are
different, i.e. the dynamical exponent $z$, the fractal dimension of
unstable sites $d_f$, and the average growth of activity $\eta$.  Note
that the boundary driven interface combines aspects of these two cases
for uniform driving.  The dynamics occurs in parallel with all
unstable sites advancing as for constant force depinning.  The
criticality is self-organized as in the extremal constant velocity
case.

The size of an avalanche, $s$, for the interface is the integrated area
during the burst resulting from pulling the end once.  It is the
difference between the final height profile after the
pull and the initial one before the pull. In the Oslo
model, $s$ is the total number of toppling events which occur after adding 
one grain of sand at the origin.  As for the case of uniform driving, the
distribution of avalanche sizes is observed numerically to obey
a scaling form \cite{christ}
\begin{equation}
P(s) \sim s^{-\tau}G(s/L^D) \quad ,
\label{dist}
\end{equation}
which is a power law with a cutoff that grows with system size $s_{co}
\sim L^D$, where $D$ is the avalanche dimension.  We now argue that
$D$ for the boundary driven interface, and hence for the Oslo sand pile model,
 is the same as for the uniformly
driven interface.  The amount of motion required
to reach the depinning transition starting from an arbitrary
configuration scales as $L^D$.  { \it Depinning occurs 
precisely when  every site
has moved at least once}.  At this instant, but not before, it
is possible to communicate information from the boundaries throughout
the system.  Thus the scaling of the
amount of motion for the SOC attractor to be
reached is independent of the boundary condition, i.e. whether we have
boundary driven SOC or extremal, uniformly driven SOC. 
It depends only on the anomalous diffusive dynamics
\cite{scaling,PaBo} of avalanches. 
This is confirmed by
numerical simulations of the Oslo model giving $D = 2.25 \pm 0.10$
\cite{christ} and extremal interface depinning giving $D = 2.23 \pm
0.03$ \cite{scaling}.  

Since the avalanche is a compact object, the size of an avalanche $s
\sim r r_{\perp}$, where $r$ is the spatial extent of the avalanche
along the internal interface coordinate and $r_{\perp}$ is the maximum
extension in the direction of growth.  If there is only one length
scale for motion in the direction of growth then $r_{\perp}\sim
r^{\chi}$, with the roughness exponent defined by the divergence of
interfacial height fluctuations with system size $w(L) \sim L^{\chi}$.
We numerically measured $H=r_{\perp}$, the maximum number of topplings
at a site vs. $r$ for $10^7$ avalanches in the Oslo model and found 
$H \sim r^{1.23 \pm 0.03}$, in excellent agreement with our prediction
$\chi=D-1$.  Since the height in the sand pile $h \sim dH/dx$, the
roughness exponent of the surface of the sand pile model is
$\chi_{pile}=\chi -1=D-2 \simeq 0.23$.

The distribution of avalanche sizes for the Oslo model, however,
is different from the uniformly driven interface due to a conservation law.
>From Eq. (4), the average size of an avalanche diverges with the
system size as $<s> \sim L^{\gamma \over \nu}$ where $\gamma = \nu D
(2-\tau)$.  Since on average each site must topple exactly once in
the critical state in
order to transport the added grain of sand to the boundary,
\begin{equation}
<s>=L \qquad {\rm and} \qquad \tau = 2 - 1/D \quad.
\label{conservation}
\end{equation}
Using $D = 2.23$, we predict $\tau=1.55$ in precise agreement with the
measured value $\tau = 1.55 \pm 0.10$ \cite{christ} for the Oslo
model.  For the boundary driven interface, this conservation law
expresses the constraint that on average each site must advance one
step when the end is dragged one step.  The value $\tau \simeq 1.55$ for the
boundary driven interface is far from $\tau \simeq 1.13$
\cite{scaling} measured when the interface is driven uniformly either
at constant force or constant velocity.  These latter systems
do not obey Eq. (5).

Christensen {\it et al.} measured the average velocity $<v>$ of tracer
grains in transit through the pile.  It was found to vanish as $<v>
\sim L^{-0.3 \pm 0.1}$ in the model.  The time unit used to
measure velocity was the number of sand grains added.  Based on a
conservation law, it was argued that the average velocity in this time
unit scales as the inverse width of the height fluctuations of the
sand pile, i.e. $<v> \sim 1/w_{pile}(L) \sim L^{-\chi_{pile}}$.  This
occurs because the model completely separates into a frozen bulk phase
where the grains never move and an active surface zone
of width $w_{pile}$ where 
transport takes place.  The collection of grains in the active zone
moves on average as an incompressible object upon addition of a
sand grain. From our result $\chi_{pile}=2-D$, we predict
$<v> \sim L^{2-D} \sim L^{-0.23}$, in reasonable agreement with the
numerical simulation results.

It is important to notice that the time unit for the Oslo model
is different from the usual time unit used for interface depinning
since each sand grain dropped results in many update steps, 
equal on average to the average duration of an avalanche $<t>$.
In accordance with Eq. (4), we propose that
the distribution of avalanche durations,
$t$,
is given by
\begin{equation}
P(t) \sim t^{-\tau_t}G(t/L^z) \quad ,
\label{distt}
\end{equation}
where the cutoff in avalanche durations $t_{co} \sim L^z$, and from
conservation of probability $z(\tau_t -1)=D (\tau -1)$.  Thus, from
Eq. (\ref{distt}) $<t>$ is diverging in the thermodynamic limit $L
\rightarrow \infty$ as $<t> \sim L^{z(2-\tau_t)} \sim L^{D(1-\tau)+z}
\sim L^{z+1-D}$.  Measuring time in the simulation in terms of update
steps, $t$, rather than sand added, we find $<v> \sim L^{1-z} \sim
L^{-0.42}$.  We numerically measured the duration of avalanches, $t$
vs.  $r$ where $t\sim r^z$, for $10^7$ avalanches in the Oslo model and
found $z=1.42 \pm 0.03$ which is the same as Leschhorn's numerically
measured value $z \simeq 1.42$ for the constant force depinning
transition \cite{leschhorn}.  Since the average duration of avalanches
is diverging with $L$, sand must be added slower and slower as
the system size is increased in order to stay at criticality.

Burridge and Knopoff \cite{bk} introduced a mechanical model for the
stick-slip dynamics of earthquake faults.  It consists of
blocks connected by harmonic springs sliding with friction.  The first
element of the block-spring chain
is connected to a driver that moves at constant velocity.  It
is referred to as the train model  \cite{maria}.  de Sousa
Vieira found that the train model exhibits
SOC unlike some other spring-block systems \cite{carlson}
where every element is connected to the driver.  The train model is
completely deterministic and contains {\it no quenched randomness nor
randomness in the initial conditions}.  The equation of motion for
the position of the $j^{th}$ block, $ {\rm U}_j$,
is
\begin{equation}
\label{train}
{\ddot U}_j 
= U_{j+1} - 2 U_j + U_{j-1} - \Phi\left({ {\dot U}_j \over \nu_c}\right)
\quad .
\end{equation}
The friction force $\Phi(0)=1$, and at finite velocity
\begin{equation}
\Phi\left({{\dot U}\over \nu_c}\right) = {{\rm sgn} \left({\dot U}\right)
\over 1 + {{\dot U} \over \nu_c}}\quad .
\end{equation}
The equation of motion is valid 
if the sum of elastic forces is greater than the static friction force;
otherwise ${\dot U}_j=0$.  The train model has
one positive Lyaponov exponent giving chaotic behavior \cite{maria2}.
The blocks in this system exhibit slip-stick dynamics with a power law
distribution of events sizes and extents.    A sum rule for the
moments of slipping events corresponding to Eq. (\ref{conservation})
holds since on average every block must move with an average velocity
equal to the pulling speed, but this motion 
takes place intermittently
in terms of bursts \cite{maria}.

We conjecture that the train model is in the same universality class
as the Oslo model and boundary driven interface depinning.  Since the
model is dissipative and exhibits SOC, it is reasonable that a
dissipative term $\dot U$ would dominate the acceleration term in
Eq. (\ref{train}) at long length and time scales.  After a slip event,
the blocks come to rest in a new configuration with a random elastic
force increment at each site required to induce a subsequent slip
event or toppling.  This is the result of the chaotic dynamics, and in
a coarse grained picture can be described by quenched random
thresholds $\Phi(0)$ in the space of position and events, which
corresponds to $\eta(x,H)$.  Such an equivalence of a deterministic
model with no embedded randomness which is chaotic with a stochastic
model also occurs between the deterministic Kuramoto-Shivashinsky
\cite{mapping} equation and the Langevin equation proposed by Kardar,
Parisi, and Zhang \cite{kpz}.  Indeed, numerical simulations of the
train model give $\tau -1 \simeq 0.6$ and $\tau_R = 1 + D(\tau -1)=D
\simeq 2.2$ \cite{maria}, which agree with the critical exponents
measured for the Oslo model and support our conjecture.

Our results imply that broad universality classes in self-organized
critical phenomena exist.  In particular we establish that the Oslo
sand pile model for transport in granular piles is in the same
universality classes as the depinning transition of an interface when
it is slowly dragged at one end through a random medium.  The mapping
unifies the mechanism giving dispersive transport with long-tailed
waiting time distributions in granular media with avalanche dynamics
of interface depinning. The purely deterministic train model of
Burridge and Knopoff describing earthquakes is proposed to be in the
same universality class.

While this manuscript was in preparation, we became aware of a work by
Cule and Hwa \cite{hwa} who demonstrate that a similar
block-spring model with quenched random spring constants
and purely dissipative dynamics is in the
universality class of interface depinning with uniform driving.  

We thank the authors of Refs. \cite{christ,hwa} for making their work
available prior to publication, and P. Bak for a number of helpful
comments on the manuscript.  This work was supported by the
U. S. Department of Energy under Contract No. DE-AC02-76-CH00016 and
DE-FE02-95ER40923.

\end{document}